\documentclass[]{spie}  

\usepackage{amsmath,amsfonts,amssymb}
\usepackage{graphicx}
\usepackage{braket}
\usepackage{xcolor,colortbl}

\usepackage{subcaption}
\usepackage[colorlinks=true, allcolors=blue]{hyperref}
\usepackage{array}
\usepackage{hhline}
\title{Assessment of Polarization Entanglement Source: Photon Counting and Correlation Measurement}

\author[a,b]{Tomáš Novák}
\author[b]{Martin Guldan}
\author[b]{Josef Vojtěch}
\author[a]{Josef Blažej}

\newcommand\blfootnote[1]{
    \begingroup
    \renewcommand\thefootnote{}\footnote{#1}
    \addtocounter{footnote}{-1}
    \endgroup
}

\affil[a]{Czech Technical University in Prague, Faculty of Nuclear Sciences and Physical Engineering, Department of Laser Physics and Photonics, Prague, 11519, Czech Republic}
\affil[b]{CESNET a.l.e., Generála Píky 430/26, 160 00 Praha 6, Czech Republic}

\begin{document} 
\maketitle
\begin{abstract}
Commercial sources of polarization entanglement at telecommunication wavelengths are already available on the market, but they lack proper certification or third-party testing. We aim to provide a comprehensive testing framework for photon counting and correlation measurements to characterize the parameters of these sources in a scalable and repeatable manner. The detection setup is included in our considerations, as the non-idealities of the components negatively affect the relevance of the measurement results. We discuss bounds for both true and false coincidences with rigorous probabilistic approach, as their ratio directly impacts the resolution of coincidence measurements and is reflected in Quantum Bit Error Rate (QBER) in the quantum telecommunication system. Quantum State Tomography (QST), polarization visibility measurements, temporal correlations measurements, and computations of other statistics are to be performed and compared at the state-of-the-art level for a three commercially available sources. Given that QST is demanding in terms of number of measurements and post-processing analysis, we discuss the relevance of determining the degree of polarization entanglement considering solely other statistics of direct measurement approach.
\end{abstract}



\section{INTRODUCTION}
\label{sec:intro}  
In the first part of this contribution, we present estimates for the expected single-detector counts and coincidence rates measured by the detection setup used. To examine this correspondence, we conduct experiments under conditions of low and high saturation, varying parameters of both the sources and detectors.

To characterize the degree of entanglement in sources of polarization-entangled single photons, we present and utilize metrics derived from QST as well as a direct measurement approach. Direct measurements are generally less experimentally demanding, particularly in post-processing, where QST metrics require constructing the state’s density matrix using a maximum-likelihood method with preliminary algebraic steps \cite{JamesQST2001}. We discuss the relationships between degree of entanglement metrics and photon counting experiments conducted on three sources of polarization entanglement at telecommunication wavelengths.    
	\blfootnote{© (2024) Society of Photo-Optical Instrumentation Engineers (SPIE). One print or electronic copy may be made for personal use only. Systematic reproduction and distribution, duplication of any material in this publication for a fee or for commercial purposes, and modification of the contents of the publication are prohibited. This is the author-prepared version. The official version is available at: \url{https://doi.org/10.1117/12.3056198}}

\section{Reference source parameters}
Suppliers of polarization entanglement sources usually state some standard parameters from which one can calculate, e.g., the expected rate of coincidence measurement, single detector counts, signal-to-noise (S/N) ratio. One of them is photon pair rate, $R_C$, which refers to the number of photon pairs produced at the source output per unit time. Heralding probability, $h$, reflects the ratio of paired photons at the source output to the total number of photons generated by the source, $R_C / R$, which is related to the measured quantities as:
\begin{equation}
    h = \frac{R_{MC} - R_{M1} R_{M2} \Delta t}{\sqrt{\eta_{T1}\eta_{T2}}\sqrt{R_{M1} R_{M2}}},  \nonumber
\end{equation}
where $R_{MC}$ is measured coincidence rate, $R_{M1}$ and $R_{M2}$ are the counts on each single-photon detector, $\eta_{T1}$ and $\eta_{T2}$ are their corresponding total detection efficiencies, and $\Delta t$ represents the time during which coincidences are accepted as true (coincidence window). A statistic to characterize source's degree of entanglement is polarization visibility of $\ket{\phi^+}$ or fidelity of a given Bell state to a measured state reconstructed by QST.

\section{Detection setup response}
    Detector’s response to illumination is strongly dependent on the level of its saturation. Saturation is reached when the average time between photon arrivals approaches or even exceeds dead-time $T_d$ \cite{Knoll1999}. At this point the total detector efficiency
    \begin{equation}
        \eta_T=\frac{\eta_Q}{1+R\eta_Q T_d} 
\label{eq:totaleq}    \end{equation}
    drops significantly and similarly for expected coincidence measurement rate 
    \begin{equation}
        R_{MC} = \frac{R_C\eta_{QA}\eta_{QB}\gamma_A\gamma_B}{(1+R\gamma_AT_d\eta_{QA})(1+R\gamma_BT_d\eta_{QB})}, \label{eq:coincM}
    \end{equation}
    where $R$ is the rate of all incident photons, $\eta_Q$ detector quantum efficiency, $\gamma$ total transmissivity over a given line and indices $A,B$ correspond to Alice and Bob. In simplest symmetric case ($\eta_{QA}=\eta_{QB}$,  $\gamma=\gamma_A=\gamma_B$) the measured coincidence rate can be estimated by $R_C \eta_T^2 \gamma^2$. 
    
    False coincidences $R_{FC}$ can be estimated similarly from accepted time delay $\Delta t$ and single detector measurements $R_M$ as $R_{FC} = R_M^2 \Delta t$ for symmetric case or $R_{FC} = R_{MA}R_{MB} \Delta t$ in case of difference in arms of Alice and Bob while coincidence windows are identical. Above-mentioned relationships demonstrate the importance of maintaining a small coincidence window to maximize the S/N.

\section{Measurement of Detection setup response }
\begin{figure}
        \begin{subfigure}[t]{0.5\textwidth}
        \centering
        \includegraphics[width=\textwidth]{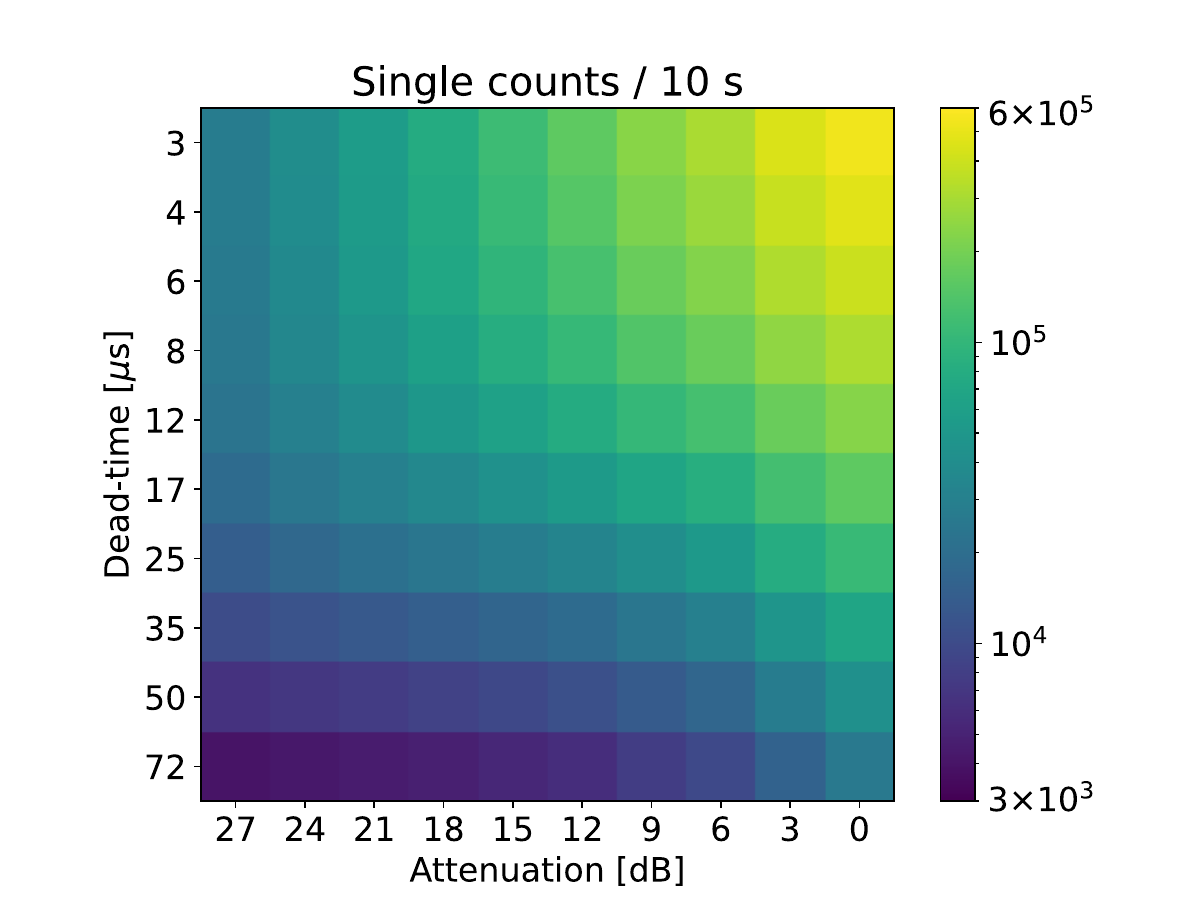}

    \end{subfigure}
    \begin{subfigure}[t]{0.5\textwidth}
        \centering
        \includegraphics[width=\textwidth]{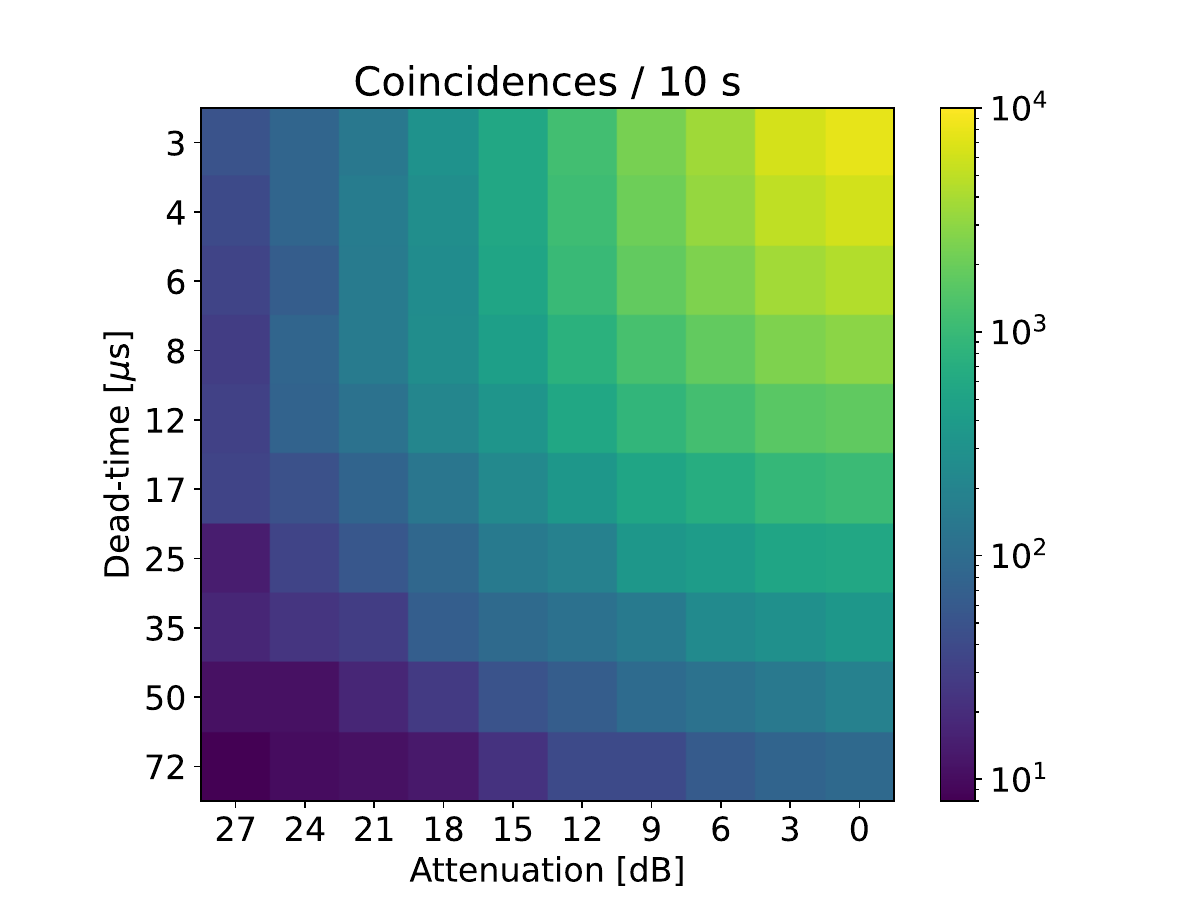}
    \end{subfigure}
    \caption{Heat maps of single detector and coincidence counts dependent on the dead-time of detectors and level of attenuation. Measurement for 100 nm broadband O-band source.}
                      \label{figrBoth}

      \end{figure}

The previous considerations were tested, with single-detector and coincidence counts plotted against dead-time and attenuation in Fig. \ref{figrBoth}. For low saturation levels, coincidence counts align closely with theoretical predictions; however, at low dead-times, single counts deviate from estimates due to the significant effects of dark counts and after-pulsing.

\begin{figure}[ht]
          \centering
          \includegraphics[width=0.5\linewidth]{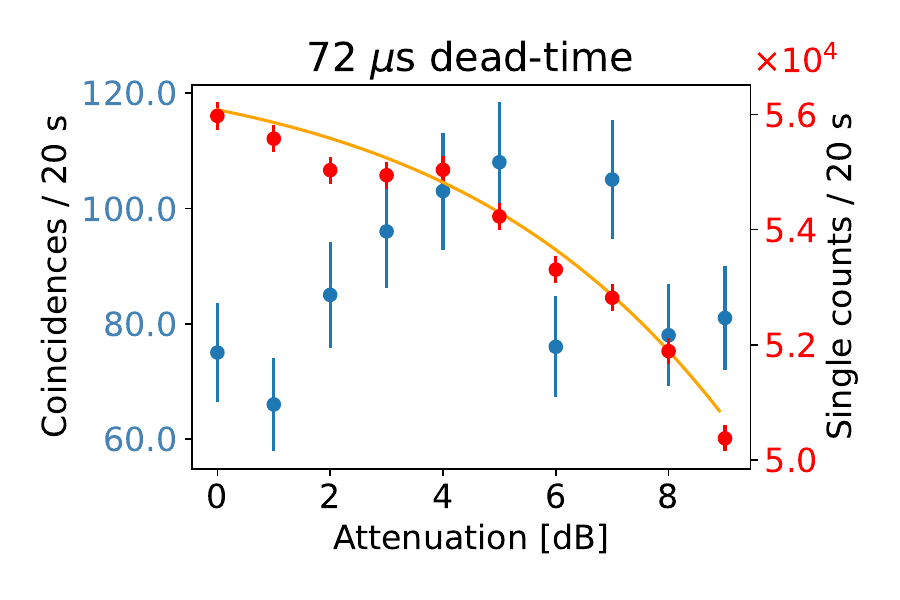}
          \caption{Coincidences and single counts as function of attenuation on both detection arms. }
          \label{figrAtt}
      \end{figure}

Figure \ref{figrAtt} shows coincidences and single counts plotted against attenuation in both detection arms under conditions of high dead-time and saturation. Notably, there is a counterintuitive increase in coincidences with increasing attenuation. This effect arises from the desynchronization of the detectors, which remain in dead-time for most of their operation. This observation highlights the importance of matching the source power to the detector counting rate to optimize coincidence rates.

\begin{figure}[ht]
          \centering
          \includegraphics[width=0.6\linewidth]{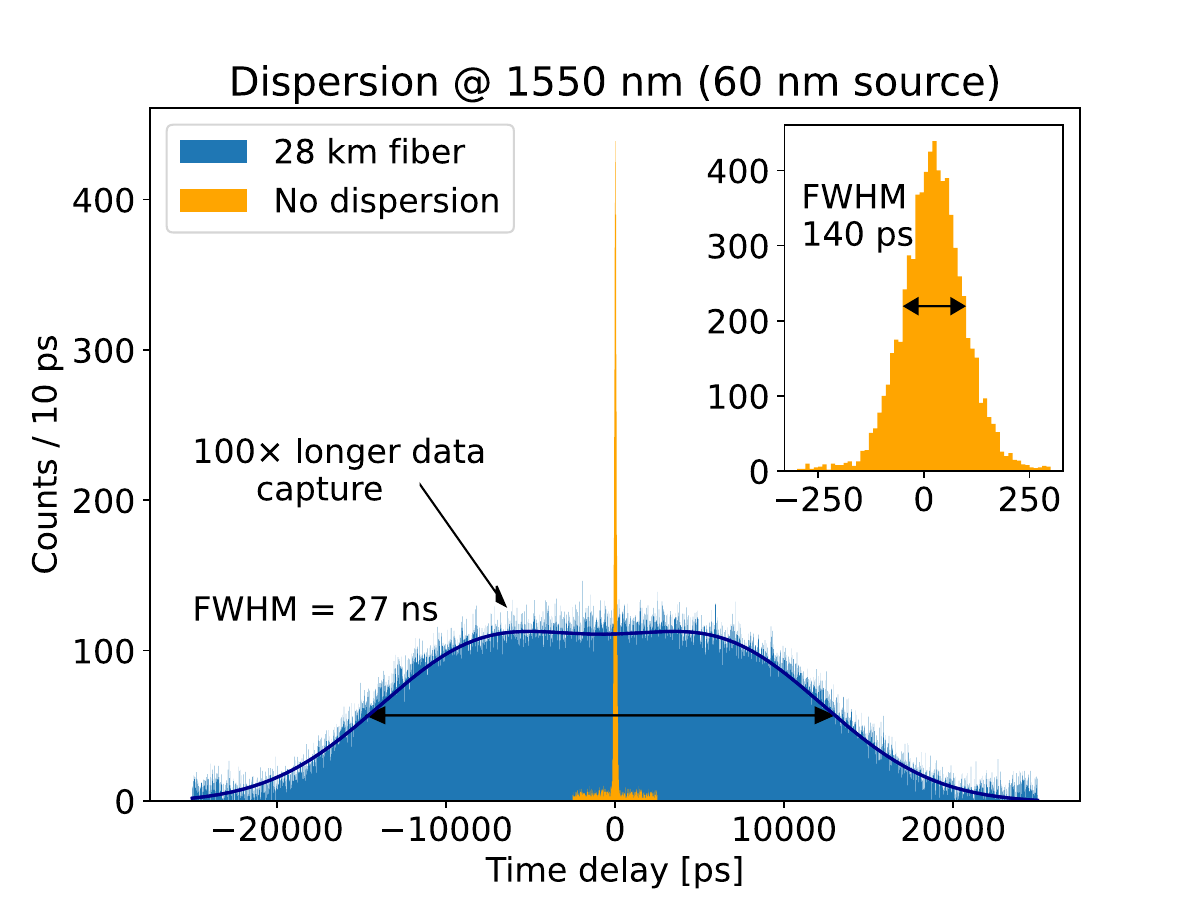}
          \caption{Correlations broadening due to dispersion in single-mode silica fibers of 60 nm broadband C-band source.}
          \label{figrDisp}
      \end{figure}

Another undesired effect is the delay broadening of paired photons in single-mode silica fibers due to chromatic dispersion, approximately 17 ps/km/nm at 1550 nm. Figure \ref{figrDisp} shows coincidence measurements of an initially narrow (limited only by detector jitter) 60 nm FWHM C-band entangled source, both at its output and after transmission through a 28 km single-mode fiber spool. The difference in FWHM between the two cases is clear, with broadening exceeding two orders of magnitude. Such broadening significantly reduces the source's suitability for telecommunication applications, as the S/N drops substantially. In the case of the 28 km fiber spool, 350 counts per bin were subtracted to improve data representation; the magnitude of this subtracted floor suggests an S/N of approximately 1/3.

\section{Metrics for degree of entanglement of direct measurement approach}
\begin{figure}[ht]
          \centering
          \includegraphics[width=0.5\linewidth]{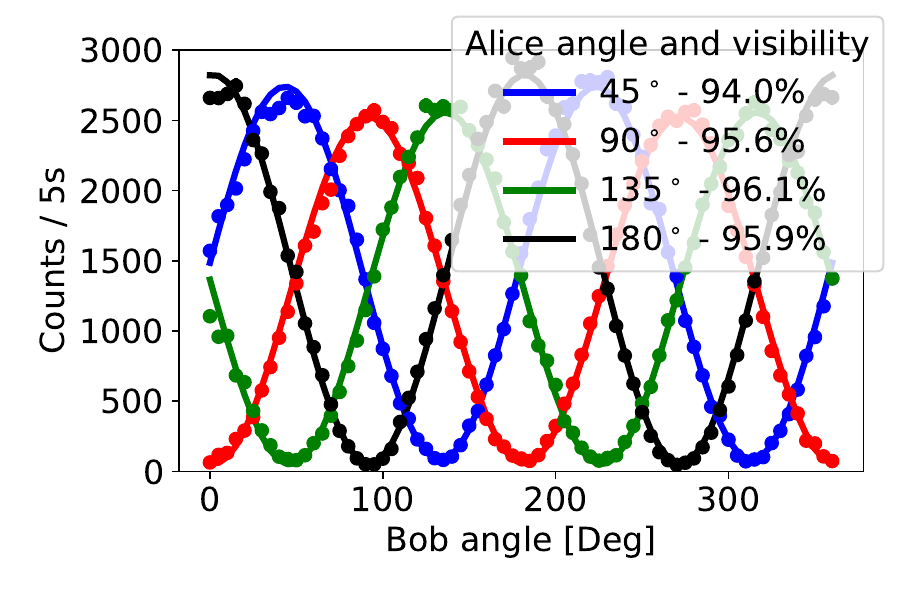}
          \caption{Measured visibility of 100 nm broadband O-band source in the Bell state $\ket{\phi^+}$ as defined by Eq. \eqref{visibility}. The angles correspond to a rotation angle of linearly polarizing filter in a given path. Orientation of $45^\circ, 90^\circ, 135^\circ, 180^\circ$ correspond to $D,H,A,V$ respectively.}
          \label{S15_visibility}
      \end{figure}

   We can derive the degree of entanglement directly from measurements, thereby reducing the complexity of both measurement and data processing. Polarization control (PC) must be adjusted to compensate for transformations in the single-mode fiber. The following metrics can be measured for any Bell state; however, in practice, the state $\ket{\Psi}$ produced by the source is typically transformed using PC to as closely approximate\cite{jin2014pulsed, shi2020stable} state $\ket{\phi^+}=1/\sqrt{2}(\ket{HH}+\ket{VV})$, meaning that the entangled photons are positively correlated in both linear ($HV, DA$) bases and anticorrelated in the circular ($RL$) base. The aforementioned polarization visibility is defined as
\begin{equation}
          V_{j} = \frac{|\braket{jj|\Psi}|^2 - |\braket{jk|\Psi}|^2}{|\braket{jj|\Psi}|^2 + |\braket{jk|\Psi}|^2},
          \label{visibility}
\end{equation}
      where the state is projected onto polarization $j\in \{H,D,R\}$ and $k\in \{V,A,L\}$ is its orthonormal counterpart, though only linear bases are typically used as in Fig. \ref{S15_visibility}.
      Additionally, $V_{HV}$ or $V_{DA}$ can be defined as the average visibility in the given basis. This metric provides a lower-effort test requiring a minimum of four measurements for $V_{HV}$ and $V_{DA}$. A high-quality source will yield $V\sim 1$ when PC is correctly configured. Visibility is related to QBER by $(1 - V)/2$, which is a critical parameter for effective quantum communication.

      
      We introduce coincidence entropies defined to be independent of the choice of a measured Bell state
          \begin{equation}
              H^{same}_j = 1 - H(p_{jj},p_{jk}), \hspace{1cm} H^{cross}_j = H(p_{jl},p_{jm}),\hspace{1cm} \mathcal{H} = \sum_j H^{same}_j + H^{cross}_j, \label{eq:bilaterentr}
          \end{equation}
          where $p_{jj}=|\braket{jj|\Psi}|^2/N_j^{same}$, $p_{jk}=|\braket{jk|\Psi}|^2/N_j^{same}$, $p_{jl}=|\braket{jl|\Psi}|^2/N_j^{cross}$, $p_{jm}=|\braket{jm|\Psi}|^2/N_j^{cross}$ are probabilities, with $j\in \{H,V,D,A\}$, $k$ being a linear polarization rotated by 90$^{\circ}$ from $j$ and $l$ and $m$ being linear polarizations rotated by 45$^{\circ}$ from $j$. The terms $N_j^{same} = |\braket{jj|\Psi}|^2 + |\braket{jk|\Psi}|^2$ and $ N_j^{cross} = |\braket{jl|\Psi}|^2 + |\braket{jm|\Psi}|^2$ and are chosen such, so the condition $p_1+p_2=1$ in the binary entropy function $H(p_1,p_2)=-p_1\log_2(p_1)-p_2\log_2(p_2)$ is satisfied. For a one-sided coincidence entropy, the inequality $\mathcal{H}\leq8$ holds. Evaluating Eq. \eqref{eq:bilaterentr} for both Alice and Bob yields $\mathcal{H}_A$ and $\mathcal{H}_B$.

Another metric for the degree of entanglement might be QBER, as it reflects the usability of the source in a quantum communication scheme. QBER is defined as the ratio of incorrect to correct coincidence counts with respect to an expected Bell state. 

\begin{figure}
        \begin{subfigure}[t]{0.5\textwidth}
        \centering
        \includegraphics[width=\textwidth]{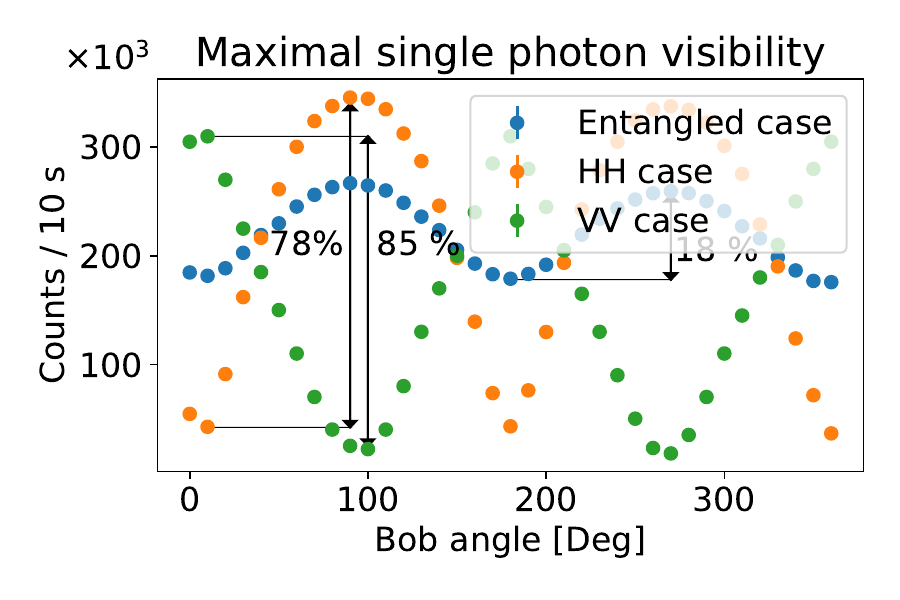}

    \end{subfigure}
    \begin{subfigure}[t]{0.5\textwidth}
        \centering
        \includegraphics[width=\textwidth]{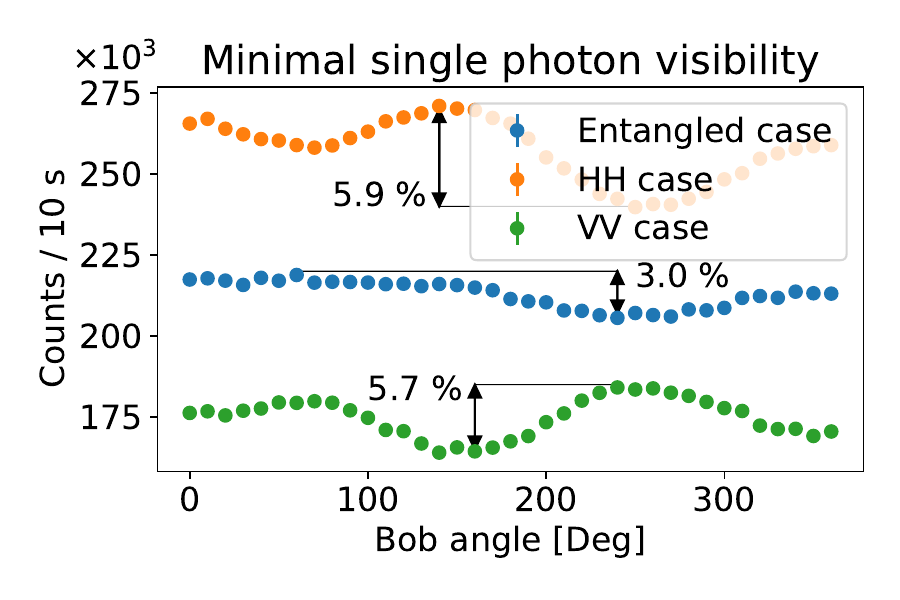}
    \end{subfigure}
    \caption{Measured maximal and minimal single-photon visibilities of 60 nm broadband C-band source. }
                      \label{OzOPt}

      \end{figure}

Single-photon visibility, $SV$, is also a candidate metric, obtained by rotating a linear polarizer in either Alice’s or Bob’s arm and determined from single-detector counts. The PC can be adjusted to achieve maximal and minimal visibility, thus introducing $SV^{max}$ and $SV^{min}$, as illustrated in Fig. \ref{OzOPt}. A plausible hypothesis is that maximal and minimal single-photon visibilities correspond to the degree of circularity of an unentangled impurity. Single-photon visibility is arguably the simplest measurement; a maximally entangled source should exhibit $SV^{max} = SV^{min} = 0$, as entangled photon pairs behave like unpolarized light for non-coincidence measurements.

\section{QST metrics for assessing the degree of entanglement}

\begin{figure}[ht]
          \centering
          \includegraphics[width=0.6\linewidth]{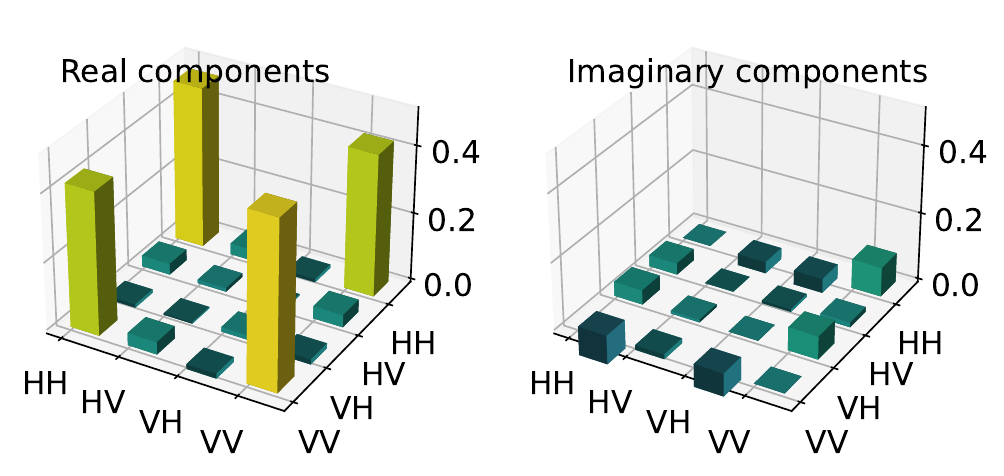}
          \caption{Reconstructed density matrix from QST of 60 nm broadband C-band source. }
          \label{figrQST}
      \end{figure}

QST is the most straightforward and reliable method for determining the quality of an entanglement source. Following mentioned method\cite{JamesQST2001} the state's density matrix, $\hat{\rho}$, is obtained from which various properties can be calculated \cite{motazedifard2021nonlocal}. Metrics like purity $p=\text{Tr}(\hat{\rho}^2)$, Von-Neumann entropy $\mathcal{S}=-\text{Tr}(\hat{\rho}\log_2 \hat{\rho})$, or Renyi 2-entropy $\Upsilon_A(\hat{\rho}) = -\ln\hat{\rho}_A^2$, where $\hat{\rho}_A$ is the reduced density matrix of Alice's subsystem (or analogously $\Upsilon_B(\hat{\rho})$ for Bob's subsystem). The density matrix is constructed from a set of coincidence measurements spanning the 16-dimensional parameter space. The density matrix depicted in the Fig. \ref{figrQST} was measured for 60 nm broadband C-band source.

\section{Results and conclusion}

From the photon counting experiments, we conclude that single-photon counts deviate from the model \eqref{eq:totaleq}, likely due to the effects of dark counts and afterpulsing at low dead-times. The behavior of coincidences in high saturation regimes, shown in Fig. \ref{figrAtt}, is also noteworthy, as the predictions of \eqref{eq:coincM} no longer hold due to detector desynchronization.

Table \ref{tab:resultss} shows that direct measurement metrics correspond well with the degree of entanglement as measured by the reference metrics obtained from QST. Notably, for the non-coincidence measurement of single-photon visibility, the difference $SV^{max}-SV^{min}$ appears to depend on the degree of entanglement rather than the individual magnitudes $SV^{max}$, $SV^{min}$. This observation suggests that the hypothesis proposing single-photon visibility originates from the circularity of an unentangled impurity is likely incorrect.

One-sided coincidence entropy, though measurement demanding, is sensitive to the degree of entanglement and is particularly well-suited for quantum communication setups, as it relies solely on coincidence measurements in the linear polarization basis. This characteristic is advantageous, as the metric can be computed from readily available counts, simplifying experimental implementation by eliminating the need for circular basis measurements, which are required in the QST approach. Additionally, the 16 defined coincidence entropy metrics for Alice and Bob may serve as monitoring statistics to detect potential adversary interference in quantum communication. A key advantage of this metric lies in the cross-basis terms in Eq. \eqref{eq:bilaterentr}, which ensure the $HV$ and $DA$ bases being mutually unbiased, this feature is not present in QBER.

In our future research, we will explore the algebraic correspondence between QST and direct measurement metrics. Additionally, we will investigate the application of one-sided coincidence entropies for automated polarization basis adjustment and for monitoring the security of quantum links in systems that use distributed polarization entanglement over single-mode fiber networks.

\begin{table}[ht]
\centering
\caption{Scalar metrics derived from QST and direct measurements for three sources of entanglement.} 
\label{tab:resultss}
\begin{tabular}
{|>{\centering\arraybackslash}m{3.cm}|
>{\centering\arraybackslash}m{1.cm}|
>{\centering\arraybackslash}m{1.cm}|
>{\centering\arraybackslash}m{1.cm}|
>{\centering\arraybackslash}m{1.cm}|
>{\centering\arraybackslash}m{1.cm}|
>{\centering\arraybackslash}m{1.cm}|
>{\centering\arraybackslash}m{1.cm}|
>{\centering\arraybackslash}m{1.cm}|
>{\centering\arraybackslash}m{1.cm}|
>{\centering\arraybackslash}m{1.cm}|}
\hline
 & \multicolumn{3}{c|}{\textbf{QST} \cellcolor{blue!25}} 
 & \multicolumn{7}{c|}{\textbf{Direct measurement} \cellcolor{green!25}} \\
\hline
\textbf{Source} & $p$ & $\mathcal{S}$ & $\Upsilon_A$ & $V_{HV}$ & $V_{DA}$ & \textbf{QBER} & $\mathcal{H}_A$ & $\mathcal{H}_B$ & $SV^{\max}$ & $SV^{\min}$ \\
\hline
\text{O-band (100 nm)} & 0.89 & 0.33 & 0.68 & 0.96 & 0.95 & 0.03 & 7.30 & 7.30 & 0.04 & 0.02 \\
\hline
\text{C-band (60 nm)} & 0.86 & 0.37 & 0.69 & 0.96 & 0.84 & 0.05 & 6.88 & 6.88 & 0.12 & 0.10 \\
\hline
\text{C-band (3 nm)} & 0.61 & 0.84 & 0.64 & 0.54 & 0.73 & 0.19 & 5.07 & 4.89 & 0.18 & 0.03 \\
\hline
\end{tabular}
\end{table}

\acknowledgments 
 
This work was supported partially by The Ministry of Education, Youth, and Sports of the Czech Republic by project EH22\_008/0004649 Quantum Engineering and Nanotechnology and partially by The Ministry of Interior of the Czech Republic through NU-CRYPT - Quantum encrypted communication with increased physical layer security (VK01030193).
\bibliography{report} 

\begin{thebibliography}{1}

\bibitem{JamesQST2001}
James, D. F.~V., Kwiat, P.~G., Munro, W.~J., and White, A.~G., ``Measurement of qubits,'' {\em Phys. Rev. A}~{\bf 64},  052312 (2001).

\bibitem{Knoll1999}
Knoll, G.~F., ``Radiation detection and measurement,'' (1999).

\bibitem{jin2014pulsed}
Jin, R.-B., Shimizu, R., Wakui, K., Fujiwara, M., Yamashita, T., Miki, S., Terai, H., Wang, Z., and Sasaki, M., ``Pulsed sagnac polarization-entangled photon source with a ppktp crystal at telecom wavelength,'' {\em Optics Express}~{\bf 22}(10),  11498--11507 (2014).

\bibitem{shi2020stable}
Shi, Y., Moe~Thar, S., Poh, H.~S., Grieve, J.~A., Kurtsiefer, C., and Ling, A., ``Stable polarization entanglement based quantum key distribution over a deployed metropolitan fiber,'' {\em Applied Physics Letters}~{\bf 117}(12),  124002 (2020).

\bibitem{motazedifard2021nonlocal}
Motazedifard, A., Madani, S.~A., Dashkasan, J.~J., and Vayaghan, N.~S., ``Nonlocal realism tests and quantum state tomography in sagnac-based type-ii polarization-entanglement spdc-source,'' {\em Heliyon}~{\bf 7}(6),  e07384 (2021).

\end{thebibliography}
\bibliographystyle{spiebib} 

\end{document}